\documentclass[prd,reprint,preprintnumbers,twocolumn]{revtex4-2}

\usepackage{amsmath,amssymb,bm,mathrsfs}
\usepackage{graphicx,color}
\usepackage{amsfonts}
\usepackage[export]{adjustbox}
\usepackage{mathtools}
\usepackage[sort&compress]{natbib} 
\usepackage{srcltx}
\usepackage{dsfont}
\usepackage{epsfig}
\usepackage{bbold}
\usepackage{rotating}
\usepackage[dvipsnames]{xcolor}
\usepackage{subfig}
\usepackage{xfrac}
\usepackage{multirow}
\usepackage{booktabs}
\usepackage[inline]{enumitem} 
\usepackage[colorlinks=true
,urlcolor=blue
,anchorcolor=bluec
,citecolor=blue
,filecolor=blue
,linkcolor=red
,menucolor=blue
,linktocpage=true
,pdfproducer=medialab
,pdfa=true
]{hyperref}
\usepackage{cleveref}
\usepackage{setspace}
\usepackage{xr}
\AtBeginDocument{%
  \heavyrulewidth=.08em
  \lightrulewidth=.05em
  \cmidrulewidth=.03em
  \belowrulesep=.65ex
  \belowbottomsep=0pt
  \aboverulesep=.4ex
  \abovetopsep=0pt
  \cmidrulesep=\doublerulesep
  \cmidrulekern=.5em
  \defaultaddspace=.5em
} 

\newcounter{qnumber}

\usepackage{tikz}

\definecolor{forestgreen}{RGB}{34,139,34}
\definecolor{nanocolor}{RGB}{0,51,51}
\definecolor{eptacolor}{RGB}{102,215,230}
\definecolor{pptacolor}{RGB}{51,102,128}
\definecolor{SMBHcolor}{RGB}{255,128,0}
\definecolor{Ourcolor}{RGB}{128,0,0}
\definecolor{myblue}{RGB}{0,0,255}
\colorlet{myblue}{White!50!myblue}

\begin{document}

\title{Using Pulsar Parameter Drifts to Detect Sub-Nanohertz Gravitational Waves} 

\author{William DeRocco}
\email{wderocco@ucsc.edu}
\affiliation{\sl Department of Physics, University of California Santa Cruz, 1156 High St., Santa Cruz, CA 95064, USA\\
and Santa Cruz Institute for Particle Physics, 1156 High St., Santa Cruz, CA 95064, USA}

\author{Jeff A. Dror}
\email{jeffdror@ufl.edu}
\affiliation{\sl Department of Physics, University of California Santa Cruz, 1156 High St., Santa Cruz, CA 95064, USA\\
and Santa Cruz Institute for Particle Physics, 1156 High St., Santa Cruz, CA 95064, USA}
\affiliation{\sl Institute for Fundamental Theory, Physics Department,
University of Florida, Gainesville, FL 32611, USA}


\begin{abstract}
Gravitational waves with frequencies below 1~nHz are notoriously difficult to detect. With periods exceeding current experimental lifetimes, they induce slow drifts in observables rather than periodic correlations. Observables with well-known intrinsic contributions provide a means to probe this regime. In this work, we demonstrate the viability of using observed pulsar timing parameters to discover such ``ultralow'' frequency gravitational waves, presenting two complementary observables for which the systematic shift induced by ultralow-frequency gravitational waves can be extracted. Using existing data for these parameters, we search the ultralow frequency regime for continuous-wave signals, finding a sensitivity near the expected prediction from inspirals of supermassive black holes. We do not see an excess in the data, setting a limit on the strain of $ 1.3 \times 10 ^{ - 12} $ at $ 450~{\rm pHz} $ with a sensitivity dropping approximately quadratically with frequency until $ 10~{\rm pHz}$. Our search method opens a new frequency range for gravitational wave detection and has profound implications for astrophysics, cosmology, and particle physics.
 \end{abstract}

\maketitle
\section{Introduction}
In the era of gravitational wave (GW) astronomy, extending our observational capacity over the frequency spectrum has become a top priority. Existing experiments cover a large range of frequencies; tests for cosmic microwave background (CMB) tensor modes~\cite{Planck:2018jri} probe the $10 ^{ - 18}~{\rm Hz} - 10 ^{ - 16} ~ {\rm Hz} $ range, existing pulsar timing array (PTA) analyses search the $ 1~{\rm nHz} - 100~{\rm nHz} $ range~\cite{NANOGrav:2020bcs,Goncharov:2021oub,Chen:2021rqp,Antoniadis:2022pcn}, and laser interferometers are already detecting GWs in the $ 10~{\rm Hz} - 1~{\rm kHz} $ range~\cite{KAGRA:2021kbb}. Future space-based interferometers, such as the Laser Interferometer Space Antenna, will cover $1~{\rm mHz} - 1~{\rm Hz} $~\cite{2017arXiv170200786A}, while underground experiments~\cite{Maggiore:2019uih,Dimopoulos:2007cj} aim to target the region between ground-based and space-based interferometry. There has been significant discussion on potential observational techniques in the $ \mu {\rm Hz} - {\rm mHz} $ range~\cite{PhysRevD.87.084009,NANOGrav:2015fwa,PhysRevLett.119.261102,Klioner:2017asb,Darling_2018,Wang:2020pmf,Fedderke:2022kxq,Fedderke:2021kuy,Blas:2021mqw,Blas:2021mpc} and above the $ {\rm kHz} $ range~\cite{Aggarwal:2020olq,Domcke:2022rgu,Berlin:2021txa} to expand our GW frequency coverage.

Despite these efforts, a significant gap in our experimental efforts remains in the spectrum between CMB polarization and PTA analyses, $ 10 ^{ - 16}~{\rm Hz} - 10 ^{ - 9}~{\rm Hz} $. This ``ultralow-frequency'' range is strongly motivated by the expectation of GWs from supermassive black hole (SMBH) binary inspirals~\cite{1980Natur.287..307B,Sesana:2013dma} (see also Refs.~\cite{Moore:2021ibq, Chang:2019mza,Neronov:2020qrl,Brandenburg:2021tmp} for examples of possible cosmic sources).

The detection of ultralow-frequency GWs is a significant experimental challenge since their period exceeds experimental timescales (e.g., up to thirty years for existing PTAs). In the context of PTAs, the effect of GWs at these low frequencies is to induce secular drifts in pulse arrival times instead of oscillatory features. Traditional pulsar timing searches fit a set of timing model parameters to the times of arrival, then subtract away this fit to search for correlations in the residuals (while simultaneously marginalizing over small deviations in the fit parameters). In this paper, we demonstrate that rather than searching for signal in the residuals, one can instead perform a search using the measured values of the fit parameters themselves. This provides a powerful alternative analysis strategy for the detection of ultralow-frequency GWs. While our analysis relies on polynomial approximations of a GW signal, hence cannot reach the sensitivity of a dedicated search using the pulse times of arrival, our analysis is intuitive, computationally-efficient, and provides an easy means of extending a search to ultralow frequencies.  

We perform a search using existing data for two complementary timing model parameters: the second derivative of the pulsar period with respect to time and the orbital decay of pulsars in binary systems. Our work builds on existing literature~\cite{10.1093/mnras/203.4.945,Kopeikin:1997rj,1999MNRAS.305..563K,2003AstL...29..241P,Kopeikin:2004hw,2009MNRAS.398.1932P,2018MNRAS.478.1670Y,2019MNRAS.489.3547K,Kumamoto:2020nas,Kikunaga:2021wwu} in multiple key ways~\footnote{Alternative approaches to detect ultralow-frequency GWs are through astrometric lensing~\cite{Gwinn:1996gv,Book:2010pf,Darling:2018hmc} and searching for CMB spectral distortions~\cite{Kite:2020uix}. While the sensitivity of these techniques is currently well below the capabilities of pulsar timing in the frequency range we consider, they form the strongest constraints for even lower frequencies.}. Firstly, we use an array of pulsars to search for GWs rather than a single, well-measured pulsar, as cross-sky correlations provide a critical method to discriminate over noise sources. Secondly, we simultaneously use information from binary pulsar orbital parameters and single pulsar parameters. As we will show, these observables are sensitive to different powers of signal frequency; detecting a signal with both methods gives critical information about the source. Finally, we apply our strategy to existing data to search for signatures of continuous GWs sourced by individual SMBH binaries. While we do not find significant evidence for GWs in our data, the results reach sensitivities to GWs within the expected range for these sources.

\begin{figure*}[t]
\begin{center} 
\includegraphics[width=16cm]{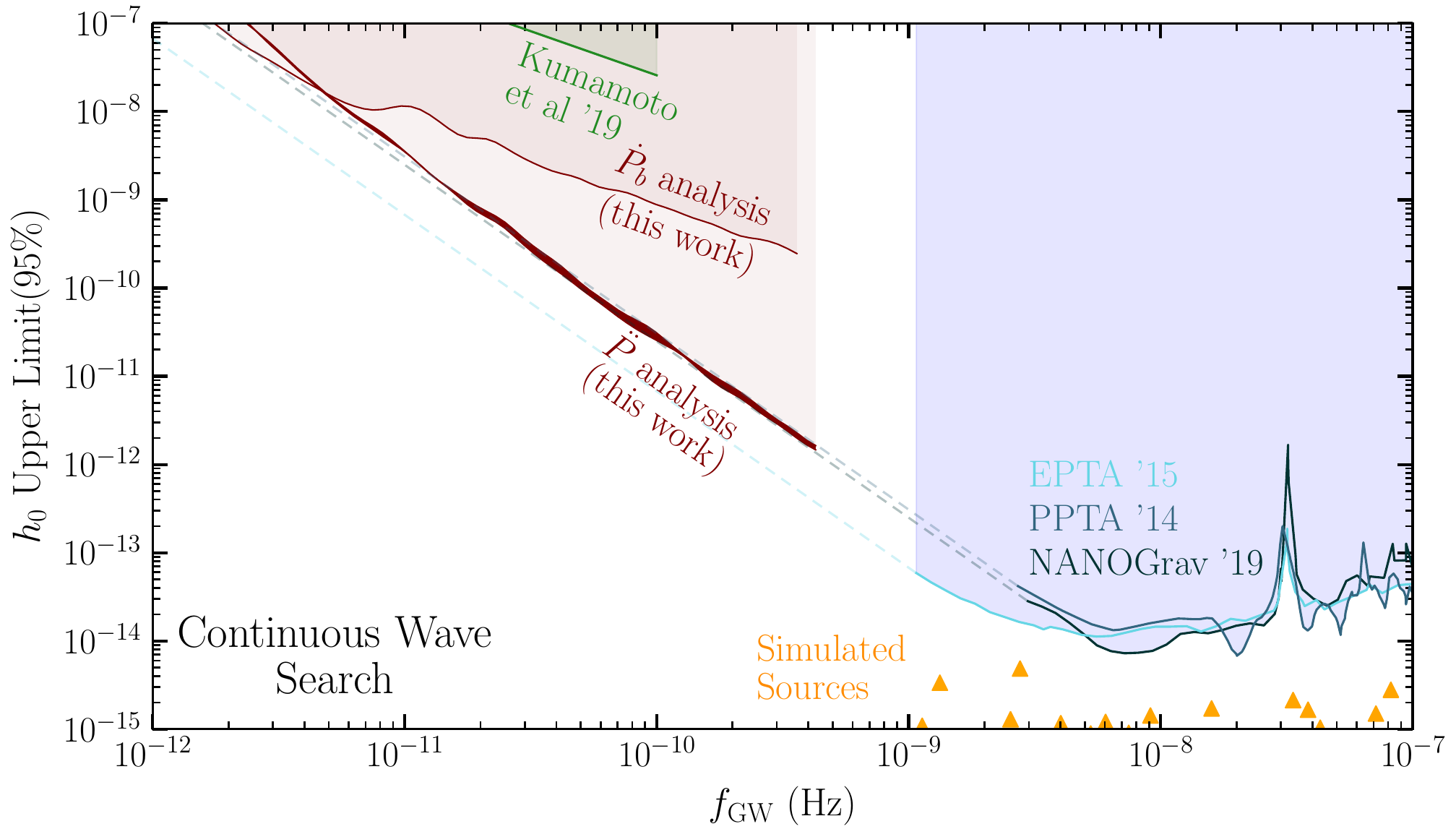} 
\end{center}
\caption{Sensitivity of pulsar timing arrays to continuous wave sources from inspirals of supermassive black hole using $ \ddot{ P} $ and $ \dot{ P} _b $ analyses ({\bf \color{Ourcolor} red}). For the $ \ddot{ P } $ analysis, we use the width of the line as an estimate of the influence of ultralow-frequency red noise (see text for detals). Constraints from traditional PTA searches shown by EPTA ({\bf \color{eptacolor} light blue})~\cite{Babak_2015}, PPTA ({\bf \color{pptacolor} blue})~\cite{10.1093/mnras/stu1717}, and NANOGrav ({\bf \color{nanocolor} dark blue})~\cite{Aggarwal_2019}. We also show results from a previous search using statistical analysis of $ \dot{ P} $~\cite{2018MNRAS.478.1670Y} ({\bf\color{forestgreen}green}). Finally, we plot the output from a simulation of SMBH mergers ({\bf \color{SMBHcolor} orange})~\cite{Kocsis:2010xa}.}
\label{fig:sensitivity}
\end{figure*}

\section{Gravitational waves in PTAs}
\label{sec:GWinPTA}
\subsection{Influence on Pulsar Timing}

Pulsar timing array experiments measure periodic radio emissions from millisecond pulsars over timescales of decades, with data taken for each pulsar approximately once every few weeks and observed for around one hour. The pulses in each observation period are folded together to produce a single time of arrival (TOA) associated with that observation, denoted $t_\text{obs}^i$ for observation $i$.

Existing PTA analyses proceed as follows: the observed TOAs $\{t_\text{obs}^i\}$ for a given pulsar are fit to a timing model $\bar{  t} ( \mathbf{\lambda})$, where we use $\mathbf{\lambda}$ to denote the set of model parameters. We assume the timing model is extensive, meaning it has sufficient parameters to encapsulate all secular drifts. (We elaborate on the influence of truncating the timing model to a non-extensive parameter set in Sec.~\ref{sec:secular}.) The output of the fitting procedure is a set of best-fit parameters $\mathbf{\hat{\lambda}}$, such that the difference between the observed TOAs and expected TOAs is minimized. These differences are the \textit{timing residuals}, $ t ^i _{\text{obs}} - \bar{ t} ^i (\mathbf{\hat{\lambda}})$. The periodic oscillations of GWs with frequencies above the inverse time span of the experiment, $f _{ {\rm GW}} \gtrsim T ^{-1}  \simeq 1~ {\rm nHz}$, induce oscillatory correlations in these timing residuals. These correlations are the target of all traditional PTA analyses.

GWs with frequencies below the inverse time span of the experiment, $f_{\text{GW}} \lesssim T ^{-1} $, manifest as low-frequency changes in the TOAs, also known as ``secular drifts.'' If the pulsar timing model is extensive, ultralow-frequency GWs are removed from the residuals by the fitting procedure. Nevertheless, one can still search for GWs through the induced biases in the best-fit parameters.

The list of model parameters used is pulsar-specific~\cite{10.1111/j.1365-2966.2006.10870.x}, though it includes the pulsar period ($ P $), its rate of change ($ \dot{ P} $), and potentially higher derivatives. If the pulsar is in a binary, the timing model will additionally include the binary period ($ P _b $) and its derivative ($ \dot{ P } _b $). The observed values of these parameters are modified in the presence of ultralow-frequency GWs. Their dominant effect can be described through an apparent relative motion between the solar system barycenter (SSB) and the pulsar. We write this effect in terms of a relative velocity (more commonly known as redshift in the literature),
\begin{align} 
v _{ {\rm GW}}( t ) &  = \sum _{ A = + , \times }F _A   ( \hat{n} ) \left(   h _A ( t ,   0 ) - h _A ( t - d _a  , {\mathbf{d}} _a  ) \right),
\label{eq:vEP}
\end{align} 
where the subscript $A = \times, +$ denotes the cross/plus polarization of the wave, and we use $ {\mathbf{d}} _a $ to denote a displacement vector from the SSB to pulsar-$ a $ in a frame with the SSB at the origin. We work in units such that the speed of light is set to unity. The $h_{+,\times} ( t , {\mathbf{x}} )$ functions describe a periodic GW source at position $ {\mathbf{x}} $. The quadrupolar nature of the wave imprints a particular pattern of contraction or expansion, captured in the ``pattern functions,''
\begin{equation} 
F  _{ A } ( \hat{n} )  \equiv \frac{ \hat{d } _a ^i \hat{d } _a ^j \hat{\bf e} _{ij} ^A ( \hat{n} ) }{ 2 ( 1 + \hat{n} \cdot \hat{d} _a ) }\,,
\end{equation} 
where $ \hat{\bf e} _{ij} $ are the polarization tensors and $ \hat{n} $ is the unit vector pointing toward the source. We provide a derivation of Eq.~\eqref{eq:vEP} to clarify its identification as a velocity in the Supplementary Material (SM), following Ref.~\cite{Maggiore:2018sht}.

Due to the strong astrophysical motivation in this frequency band, we search for a signal from an individual SMBH binary. The form of $ h _A ( t , {\mathbf{x}} ) $ for this source is well-known and is given approximately as a sinusoid with frequency $f_\text{GW}$ and amplitude set by the dimensionless strain $h_0$. The full form and a discussion of the associated orbital parameters are presented in the SM.

\subsection{GWs as Secular Drifts}
\label{sec:secular}
The fitting procedure captures induced accelerations and jerks as systematic shifts in the observed parameter values; we explicitly include the subscript ``obs'' to remind the reader that the best-fit parameters are \textit{not} equal to the true physical parameters. For the pulsar period ($ P $) and binary period ($ P _b $), the observed values are close to the fundamental values, and we do not include this subscript on their symbols. The observed parameters contain known contributions that can be generically broken into three main classes:
\begin{enumerate}[label=(\arabic*),itemsep=0ex,topsep=5pt]
\item \textit{intrinsic} contributions, which are due to physical effects within the pulsar system such as electromagnetic or gravitational radiation liberating energy from the system, 
\item \textit{observational} contributions, which are effects induced due to relative motion between the SSB and pulsar, and
\item \textit{galactic} contributions, which are induced by the Milky Way potential.
\end{enumerate}
The timing model parameters can also be biased if pulsars have an unknown wide binary companion, are in a globular cluster, or carry mischaracterized ultralow-frequency red noise. We search for such effects in the SM. 

We now consider the known contributions to the pulsar timing model parameters for models of millisecond pulsars, starting with the observed pulsar spin-down rate, 
\begin{equation} 
\frac{ \dot{ P} _{\text{obs}}}{  P  }    = \frac{ \dot{ P }_{\rm int}}{P} -  \frac{ v _{ \perp } ^2 }{ d _a  }  - a _{ \text{MW}}  - a _{\text{GW}}\,. \label{eq:Pd}
\end{equation}
The first contribution on the right-hand side is the intrinsic spin-down of the pulsar due to electromagnetic radiation. The second term is a kinematic term arising from the motion of the pulsar (the ``Shklovskii effect''~\cite{1970SvA....13..562S}) proportional to the relative motion of the pulsar perpendicular to the line of sight ($v _\perp $). The third term is a relative acceleration induced by the Milky Way potential. The fourth term is a relative acceleration induced by a passing GW (which we aim to observe and can be calculated from Eq.~\eqref{eq:vEP}).

While the effect of GWs is contained within $ \dot{ P } _{ {\rm obs}} $, it is necessary to subtract off the intrinsic, kinematic, and galactic contributions to $\dot{P}_\text{obs}$ to extract it. Despite the observed value of this parameter being precisely measured (typical uncertainties on $ \dot{ P} _{ {\rm obs}} / P $ reach $ 10 ^{ - 24}~{\rm sec} ^{-1} $), extracting the GW contribution is not feasible due to the inherent uncertainty in the intrinsic spin-down contribution. Predicting $ \dot{ P } _{ {\rm int}} $ of a millisecond pulsar, whose value is of order a million times larger than the uncertainty on $ \dot{ P} _{ {\rm obs}} $, requires modeling the complex magnetic structure surrounding the pulsar, a procedure subject to large systematic uncertainties. These uncertainties make any extraction of a GW signal on a pulsar-by-pulsar basis impossible. Since GWs have a characteristic pattern across the sky that is uncorrelated with $ \dot{ P} _{ {\rm int}} $, GWs can still be extracted from $ \dot{ P} _{ {\rm obs}} $ statistically, as proposed and studied in Refs.~\cite{2018MNRAS.478.1670Y,2019MNRAS.489.3547K,Kumamoto:2020nas,Kikunaga:2021wwu}.

Statistical measurements of GWs using $ \dot{ P} $ are limited since we have precise measurements of approximately one hundred stable millisecond pulsars. A more sensitive approach is to consider model parameters with relatively small known contributions that can be precisely estimated. Two such parameters are the derivative of the binary period $\dot{P}_b$ and the second derivative of the pulsar spin period $\ddot{P}$. The contributions to their observed values are
\begin{align}
\frac{ \dot{ P}_{b,{\rm obs}}}{  P_b  }   & = \frac{ \dot{ P } _{b, {\rm int}}}{P_b} -  \frac{ v _{ \perp } ^2 }{ d _a  }  - a _{ \text{MW}} - a _{ \text{GW}}\,, \label{eq:Pbd}\\ 
\frac{ \ddot{ P }_{\text{obs}}}{P} & =  j _{\text{GW}}\,.   \label{eq:Pdd}
\end{align}
Eq.~\eqref{eq:Pbd} is identical in structure to Eq.~\eqref{eq:Pd}, but the intrinsic contribution is now driven by gravitational radiation emitted by the binary system. Measurements of $ \dot{ P} _{ b, {\rm obs}} /P _{ b} $ can reach uncertainties similar to those on $ \dot{ P } _{ {\rm obs}}/ P $ but has a critical difference: the value of $ \dot{ P } _{b, {\rm int}}/P _b   $ is predictable once the properties of the binary components are determined. The dominant uncertainty in isolating Eq.~\eqref{eq:Pbd} for a passing GW for many of the most sensitive pulsars is in estimating the Shklovskii term. This requires an independent measurement of $ v _\perp $ and $ d _a  $ and can be achieved through very-long-baseline interferometry, astrometry, and pulsar timing. The Milky Way potential is typically an insignificant contribution to Eq.~\eqref{eq:Pbd}, though it can be modeled. With all these contributions estimated, we can extract GW-induced acceleration. 

In the case of $\ddot{P}$, there are a set of corrections analogous to those in Eq.~\eqref{eq:Pbd}. However, the current uncertainty on $ \ddot{ P } _{ {\rm obs}} $ is too large to detect their values for old millisecond pulsars. Models of magnetic dipole braking suggest the intrinsic contribution to Eq.~\eqref{eq:Pdd} should be of order $ ( \dot{ P} /P) ^2  $~\cite{2012hpa..book.....L}, which is typically on the order of $  10 ^{ - 35}~{\rm sec} ^{-2}  $, much below typical uncertainties on the observed value and below the gravitational wave strength we wish to target. Furthermore, since kinematic and galactic contributions to $ \ddot{ P } _{ {\rm obs}}$ are suppressed due to the non-relativistic nature of the galaxy, they can be neglected. We calculate the form of corrections to Eq.~\eqref{eq:Pdd} and estimate their size in the SM using the formalism presented in Ref.~\cite{10.1093/mnras/sty1202}. 

The sensitivity of PTA analyses to ultralow-frequency GWs is not limited to the timing model parameters presented here. Firstly, one could consider higher-order derivatives of $ P $ or $ P _b $. Similar to $ \ddot{ P} $, these will have negligible intrinsic, observational, and galactic effects. While searches for higher pulsar derivatives will not improve the signal sensitivity, their relative contributions can be used to learn about the frequencies present within a passing GW~\footnote{Additionally, it has been suggested in Ref.~\cite{Kopeikin:1997rj} that the secular drift in the projected semimajor axis ($ x  $) of a binary pulsar could be used for ultralow frequency GW detection. Since $ \dot{x} / x $ has a Shklovskii contribution identical to that of $ \dot{ P} _b/ P _b  $, measurements of GWs using both $ \dot{x} $ and $ P _b $ are highly correlated and would not substantially boost experimental sensitivity.}.

So far, we have assumed the pulsar timing model includes $ \ddot{ P} $ (and its derivatives) such that the entire signal of ultralow-frequency GWs is fit into the timing model parameters. An alternative approach is to only incorporate parameters in the timing model fit that are expected to have significant non-GW contributions. In our context, this would correspond to fitting for $ \dot{ P} $ and $ \dot{ P} _b $ but not fitting for $ \ddot{ P} $. With this approach, the residuals can be used to search for GWs. In fact, theoretical projections for PTA capabilities occasionally incorporate this effect in their sensitivity estimates (see, e.g., Refs.~\cite{2015CQGra..32e5004M,Hazboun:2019vhv}), even though PTA collaborations have refrained from extending their existing curves into the ultralow-frequency regime. However, our approach has a few major advantages over the conventional analysis strategy. Firstly, timing model parameters for which additional internal contributions can be independently measured (e.g. the three non-GW contributions to $ \dot{ P } _{b,\text{obs}} $ in Eq.~\eqref{eq:Pbd}) allow us to exploit our knowledge of their values to boost sensitivity, a procedure that has no direct analog in a conventional residual search. This ability to subtract known contributions will become even more critical as PTA sensitivity improves to the point where experiments will detect kinematic corrections to $ \ddot{ P} _{ {\rm obs}} $. Secondly, our analysis can be extended to search for stochastic signals in the ultralow-frequency regime~\cite{DeRocco:2023qae}. This region is difficult to study via residuals alone since the residual correlator becomes non-stationary. Finally, studying the timing model parameters offers a significant computational benefit since it does not require simultaneously analyzing the residuals for all the pulsars in the array; each timing model parameter can be determined individually, and their biases can be subsequently used to search for GWs. 

\section{Dataset Description}
\label{sec:dataset}
We use different sets of pulsars for the two parameters of interest.  For the $\dot{P}_b$ search, we use a set of $ 14 $ binary pulsars compiled in Ref.~\cite{2021ApJ...907L..26C} to detect the Milky Way potential. (See Ref.~\cite{Phillips:2020xmf} for a similar analysis.) These pulsars were selected as they possess well-estimated intrinsic and Shklovskii contributions to the observed parameter, $ \dot{ P } _{ b , {\rm obs}}$ (first and second terms on the right-hand side of Eq.~\eqref{eq:Pbd}). We must additionally include an estimate of the contribution from the Milky Way potential (the third term on the RHS of Eq.~\eqref{eq:Pbd}), which we calculate using the \texttt{MWPotential2014} model implemented in the \texttt{galpy} Python package~\cite{Bovy_2015}. We take a 20\% uncertainty on the value for every pulsar, which is roughly the order of the uncertainties on galactic fit parameters in \texttt{MWPotential2014}~\cite{Bovy_2015}. We then subtract these three contributions from $ \dot{ P } _{b, {\rm obs}} / P _{ b }$, estimating the line-of-sight acceleration $ a _{ {\rm GW}} $ for each pulsar. A summary of all the pulsars used in both analyses, including the size of the intrinsic, Shklovskii, and MW contributions, can be found in the SM, which includes Refs.~\cite{Reardon_2021,Fonseca_2016,10.1093/mnras/stv2395,10.1093/mnras/stw483,Fonseca_2014,Alam_2020,doi:10.1126/science.1132305,10.1093/mnras/sty2905,10.1111/j.1365-2966.2012.21253.x,Liu_2020,Cognard_2017,Kaplan:2016ymq,Liu:2019iuh,Bilous_2015,1994ApJ...428..713K}. 

Measurements of $ \ddot{ P}  $ are not published by the pulsar timing collaborations for most pulsars. Instead, we use a study carried out in Ref.~\cite{Liu:2019iuh}, which searched for evidence of jerk within $  49 $ pulsars from EPTA~\cite{10.1093/mnras/stw483} and PPTA~\cite{10.1093/mnras/stv2395} data (with additional timing data from Ref.~\cite{1994ApJ...428..713K}). Of the 49 pulsars, 3 are unsuitable for searching for ultralow-frequency GWs, and we omit these in our analysis. 

The measured values of $\ddot{P}_\text{obs}$ provided by Ref.~\cite{Liu:2019iuh} include the effects of dispersion measure (DM) variation and red noise. DM variation was assumed to give rise to a Kolmogorov turbulence spectrum with an amplitude extracted by earlier fits performed by the pulsar timing collaborations. The red noise spectrum was modeled as a broken power-law with a spectral index and amplitude. Red noise was included by adding $ A _k \sin ( 2\pi f _k /T ) + B _k \cos ( 2\pi f _k /T ) $ to the residuals, where $ f _k = k / T  $ with $ k = 1 ,2, ... $ and $ A _k $ and $ B _k $ are randomly sampled from the power spectrum. By construction, this method only incorporates red noise above $ 1/ T $ and does not account for potential ultralow-frequency red noise, which influences the timing model similar to gravitational waves but is uncorrelated among the pulsars. If left as a free parameter, the ultralow-frequency red noise contribution would eliminate any prospect for gravitational wave detection. Instead, we estimate the induced variance in the fit parameters, assuming the spectrum persists as a broken power-law below $ 1/ T $ and using the fit parameters from EPTA~\cite{10.1093/mnras/stw483} and PPTA~\cite{10.1093/mnras/stv2395}, as described in the SM. We add the square root of this variance in quadrature with the uncertainty on $ \ddot{ P} _{ {\rm obs}} $. Due to uncertainty in these estimates, we perform a second analysis, where we inflate the value of the red noise estimate by an order of magnitude. Even with these inflated values, the effect of ultralow-frequency red noise does not appreciably alter the limits (see Fig.~\ref{fig:sensitivity}). For completeness, we also estimated the influence of ultralow-frequency red noise on $ \dot{ P} _b $ with the formalism introduced in the SM. We find it is always negligible with respect to the current uncertainties on $ \dot{ P} _{ b , {\rm obs}} $. 

Given that all non-GW contributions are either already accounted for in the measured $\ddot{P}$ values (as is the case for DM and red noise) or are significantly below the current uncertainties on these values (as is the case for the intrinsic, kinematic, and Galactic contributions), we produce a set of line-of-sight jerk estimates by taking $j_{\text{GW}} = \ddot{P}_\text{obs}/P$ for these 46 pulsars.

Note that at this time, the published datasets we use are 6-7 years outdated; our limits would be improved by an updated analysis. Furthermore, adding existing NANOGrav data to the analysis would substantially improve the sensitivity.

\section{Results and Discussion}
\label{sec:results}
We conduct two separate analyses using the $ a _{ {\rm GW}} $ and $ j _{ {\rm GW}} $ datasets, performing log-likelihood ratio tests for the presence of a gravitational wave signal within each dataset (see SM for details, which uses Refs.~\cite{1987GReGr..19.1101W,Algeri_2020}). We do not find significant evidence for a continuous wave signal using the datasets described in the previous section. Consequently, we set limits on $ h _0 $ as a function of $ f _{ {\rm GW}} $ and find the results shown in Fig.~\ref{fig:sensitivity} (red) for the $ \dot{ P } _b $ and $ \ddot{ P} $ analyses. For comparison, we also show limits for continuous wave sources from EPTA (light blue)~\cite{Babak_2015}, PPTA (blue)~\cite{10.1093/mnras/stu1717}, and NANOGrav (dark blue)~\cite{Aggarwal_2019}, as well as previous limits set using all-sky correlations of $\dot{P}$~\cite{2018MNRAS.478.1670Y} in green. We extend the PTA curves according to the expected scaling behavior of conventional analyses in the ultralow-frequency regime~\cite{Hazboun:2019vhv}, however, we note that none of the collaborations have published limits in this frequency range. To emphasize the power of exploring the ultralow-frequency regime for SMBH inspirals, we also show the results of a simulation of individual SMBH sources~\cite{Kocsis:2010xa}. The low-frequency cutoff of simulated sources depends sensitively on the astrophysical assumptions about SMBH binaries at separations near 1 pc; the $ \dot{ P} _b $ and $ \ddot{ P} $ analyses are a novel probe into physics on these scales.

For frequencies $ 10~{\rm pHz} \lesssim f _{ {\rm GW}} \lesssim  450~{\rm pHz} $, we find the sensitivity
\begin{equation}
 h_0 \simeq \left\{    \begin{array}{lr}\displaystyle
         1.8 \times 10^{-10} \left(\frac{\text{400 pHz}}{f_\text{GW}}\right) & (\text{$ \dot{ P} _b $ analysis})\\[8pt]
 \displaystyle                         1.6 \times  10^{-12} \left(\frac{\text{400 pHz}}{f_\text{GW}}\right)^2 & (\text{$ \ddot{ P}$ analysis})
    \end{array}\right.\,.
    \label{eq:fscaling}
\end{equation}
The scaling of the limits with frequency can be understood by observing that, in this range, $ a _{ {\rm GW}} \propto    h _0  f _{ {\rm GW}} $ and $ j _{ {\rm GW}}  \propto    h _0 f _{ {\rm GW}} ^2 $. The scaling of the $ \ddot{ P} $-analysis agrees with the scaling of projections for searches using the residuals directly at ultralow frequencies when the timing model excludes $ \ddot{ P} $~\cite{2015CQGra..32e5004M,Hazboun:2019vhv}. (This is to be expected since, in such an analysis, the unfit $ \ddot{ P} $ signal is contained in the residuals.) The $ \ddot{ P } $ analysis reaches a smaller strain for $ f  _{ {\rm GW}}\gtrsim  3.5~ {\rm pHz} $, while the $ \dot{ P} _b $ analysis reaches a smaller strain for $ f _{ {\rm GW}} \lesssim 3.5~ {\rm pHz} $. We note that a simultaneous observation of GWs using both $ \dot{ P} _b $ and $ \ddot{ P} $ would break the degeneracy of $h_0$ and $f_\text{GW}$. As such, including both may prove a critical tool for upcoming analyses. 

The behavior of the limits changes outside this frequency range. For $ f _{ {\rm GW}} \lesssim 10 ~ {\rm pHz} $, the GW frequency is below $  d _a ^{-1} $ for the most sensitive pulsars such that the GW influences both the SSB and the pulsar similarly. This causes a partial cancellation between the two terms in Eq.~\eqref{eq:vEP}. This only holds at leading order in $ f _{ {\rm GW}} $ such that the sensitivity in each case falls off as an additional power of frequency, i.e., $ a _{ {\rm GW}} \propto h _0 d _a f _{ {\rm GW}} ^2 $ and $ j _{ {\rm GW}} \propto h _0 d _a f _{ {\rm GW}} ^3  $ in this regime. For frequencies near or above $ T ^{-1} $, $ \dot{P} _b $ and $ \ddot{ P} $ are not well approximated by time derivatives of $ v _{ {\rm GW}} $. For this reason, we cut off our analysis at $ f _{ {\rm GW}} = ( 4T ) ^{-1}  $ with $ T =22$ (17.7) yr, corresponding to the longest pulsar observation time in the $\dot{P}_b$ ($\ddot{P}$) dataset.

The sensitivity achieved by searching for drifts in the timing model parameters is competitive with current PTA strategies near 1~nHz. If the gravitational wave signal observed by the pulsar timing collaborations~\cite{NANOGrav:2023gor,EPTA:2023fyk,Reardon:2023gzh} is from SMBH inspirals, then a corresponding signal is expected in the sub-nHz band. Consequently, correlated timing model drifts should appear in the near future and may be the key to uncovering the physics of SMBH binaries at separations near 1~pc. Such correlations may already be detectable by NANOGrav or with more current EPTA and PPTA observations, strongly motivating the case for the collaborations to perform measurements of $\ddot{P}$ with their existing data. 

While the focus of our work has been on continuous wave sources, our study can be extended to search for a stochastic ultralow-frequency GW background~\cite{DeRocco:2023qae}. Apart from the signal induced by SMBH inspirals, a stochastic signal in this frequency range could also be an indication of a turbulent QCD phase transition~\cite{Neronov:2020qrl,Brandenburg:2021tmp} or a consequence of new global~\cite{Chang:2019mza} or gauge symmetries~\cite{Kitajima:2022lre}. If a stochastic background were present, it could, in principle, be distinguished from a continuous source through the different correlations in the timing model parameters. We leave the viability of discriminating continuous and stochastic sources for future work. By studying biases in the timing model parameters, we open a new frequency range for exploration for PTA analyses, with profound implications for astrophysics, cosmology, and particle physics.

\vspace{0.2cm}
\noindent {\it Acknowledgements.}
We thank Sukanya Chakrabarti and Nihan Pol for useful discussions early on during this work. We also thank the anonymous referee for bringing the importance of ultralow-frequency red noise to our attention and Xiao-Jin Liu for correspondence on Ref.~\cite{Liu:2019iuh}. The research of JD is supported in part by NSF CAREER grant PHY-1915852 and in part by the U.S. Department of Energy grant number DE-SC0023093. WD is supported in part by Department of Energy grant number DE-SC0010107. Part of this work was performed at the Aspen Center for Physics, which is supported by National Science Foundation grant PHY-1607611.

\bibliographystyle{JHEP}
\bibliography{refs}

\clearpage
\onecolumngrid
\begin{center}
   \textbf{\large SUPPLEMENTARY MATERIAL \\[.2cm] ``GW Detection Below 1~nHz with Pulsar Timing Model Drifts''}\\[.2cm]
  \vspace{0.05in}
  {Jeff A. Dror, William DeRocco}
\end{center}
\setcounter{equation}{0}
\setcounter{figure}{0}
\setcounter{table}{0}
\setcounter{page}{1}
\setcounter{section}{0}

\makeatletter
\renewcommand{\thesection}{S-\Roman{section}}
\renewcommand{\theequation}{S-\arabic{equation}}
\renewcommand{\thefigure}{S-\arabic{figure}}
\renewcommand{\thetable}{S-\arabic{table}}
\renewcommand{\bibnumfmt}[1]{[S#1]}
\renewcommand{\citenumfont}[1]{#1}

\section{Methods}
\label{app:methods}
In this section, we give further details of the datasets and describe the methods used to carry out the analysis.

\subsection{SMBH Binary Signal}
\label{app:SMBHform}
Eq.~\eqref{eq:vEP} applies to any GW source $ h _A ( t , {\mathbf{x}} ) $. In light of the strong observational potential provided by supermassive black hole (SMBH) inspirals, we take the continuous-wave signal induced by an {\em individual}, circular, SMBH binary as our fiducial target. An SMBH binary source is described by eight parameters: two angles corresponding to the source direction, two angles ($ i $ and $ \psi $) corresponding to the normal vector of the binary, the chirp mass ($ {\cal M} $) and luminosity distance ($d _L $), an approximately-constant angular frequency of the binary ($ \omega _0 $), and its orbital phase offset ($ \Phi _0 $). With these parameters specified, the resulting $ h _A ( t , {\mathbf{x}} ) $ has a known form. Defining the amplitude, $h_0$, as
\begin{align} 
h _0 & \equiv \frac{ 2( G {\cal M} ) ^{ 5/3} }{ d _L }  \omega _0   ^{ 2/3} \,,\\ 
& \simeq  10 ^{ - 15} \left[ \frac{ {\cal M} }{ 10 ^{ 10 } M _{\odot} } \right] ^{ 5/3} \left[ \frac{ \omega _0/ \pi  }{ 1~{\rm nHz} } \right] ^{ 2/3} \left[ \frac{ 500~{\rm Mpc}}{d _L } \right] \,, \label{eq:h0}
\end{align}
where $ G $ denotes Newton's constant, the GW signal is~\cite{1987GReGr..19.1101W},
\begin{align} 
\hspace{-8pt}\left[ \begin{array}{c} 
h _+  \\  
 h _\times 
\end{array} \right]\hspace{-2pt} = h _0 \left[ \begin{array}{c@{\hskip 5pt}c} 
\cos 2 \psi   &  \sin 2 \psi  \\  
\sin 2 \psi  & - \cos  2 \psi 
\end{array} \right] \hspace{-2pt}\left[ \begin{array}{c} 
( 1\hspace{-2pt}+ \cos ^2 i ) \sin 2 \Phi  \\  
 2 \cos i \cos 2 \Phi  
\end{array} \right]\hspace{-2pt},
\label{eq:hcont}
\end{align} 
where $ \Phi = \Phi _0 + \omega _0  t $. The frequency of GWs is related to the orbital angular frequency, $ f _{ {\rm GW}}   = \omega _0/ \pi    $. For ultralow-frequency GWs, $f_\text{GW} t \ll 1  $ (any time of interest is less than the total observation time), such that we can expand the cosine and sine of $\Phi(t)$ resulting in an instantaneous relative velocity ($v_\text{GW} \sim h _0 $), acceleration ($a_{\text{GW}}  \sim \omega _0 h _0 $), and jerk ($j_\text{GW} \sim \omega ^2 _0 h  _0 $).

\subsection{Analysis Datasets and Validation}
\label{app:datasets}
\begin{figure}[t]
\begin{center} 
\begin{tikzpicture} 
\node at (2,0) {\includegraphics[height=10cm]{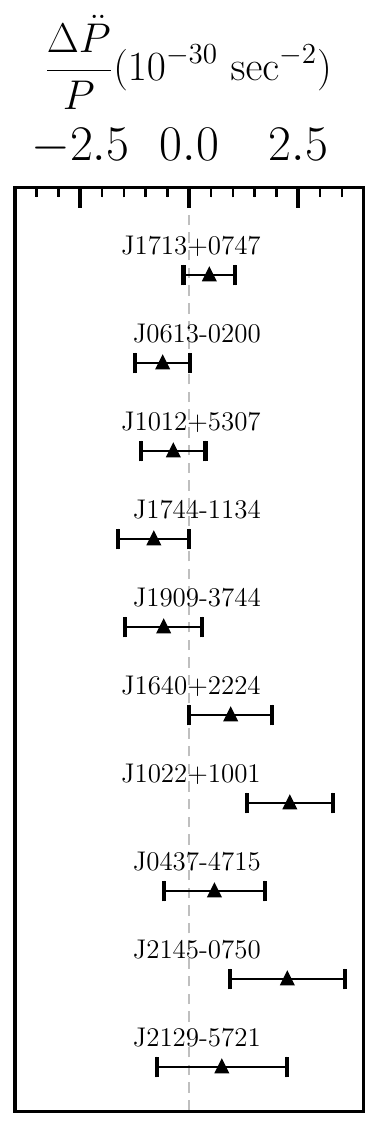}};
\node at (-2.2,-0.) {\includegraphics[height=10.cm]{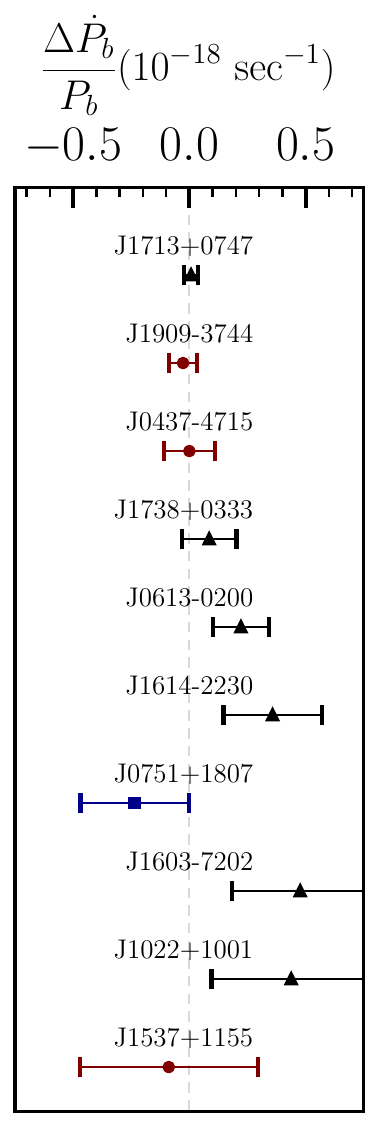}};
\end{tikzpicture}
\end{center}
\vspace{-0.75cm}
\caption{Difference between the observed and expected $ \dot{ P} _b / P _b $ {\bf (Left)} and $ \ddot{ P}/ P $ ({\bf Right}), neglecting background gravitational waves, for the ten most sensitive pulsars in each dataset. Shown are the $ 1 $-sigma error bars assuming correlations between errors are negligible. The marker type denotes the dominant contribution to the uncertainty (always the observed value for $ \ddot{ P} $).} 
\label{fig:pulsars}
\end{figure}
In our analysis, we used two different datasets to measure continuous source GWs. In this section, we provide the data for each analysis. Furthermore, we check our data for unmodeled contributions to $ \dot{ P} _b $ or $ \ddot{ P} $. As discussed in the main text, these may arise from unidentified wide binary companions to the pulsar or mischaracterized ultralow-frequency red noise. We emphasize that while our datasets are well-suited to place a limit on the presence of gravitational waves, any claim of a positive signal would require additional scrutiny to ensure the signal is arising as a correlated signal within the dataset.
 
In Tab.~\ref{tab:Pbdot}, we show the set of binary pulsars used to carry out the $ \dot{ P} _b $ analysis. We chose this set since each pulsar's proper motion and distance were independently measured. These are needed to measure the Shklovskii term in Eq.~\eqref{eq:Pbd} and hence isolate for the GW contribution to $ \dot{ P } _{ b , {\rm int}} / P _b $. In addition to the coordinates and distance for each pulsar, we list $ \dot{ P} _{ b , {\rm obs}} / P _b $ and its constituent contributions, with $ \Delta \dot{ P} _b / P _b  $ denoting the estimate of a potential background GW contribution using Eq.~\eqref{eq:Pbd}. When possible, the intrinsic values were extracted from the cited reference. For PSR J0613-0200, PSR J1614-2230, and PSR J1713+0747, we used the quadrupole radiation formula to estimate the intrinsic value ourselves. A summary of the ten most sensitive pulsars is presented in Fig.~\ref{fig:pulsars} ({\bf Left}).

To validate the dataset, we search for individual outliers. By removing any outliers, we ensure any prospective signal arises primarily from the correlations within the dataset, as opposed to a single pulsar driving the fit. This process weakens our limit on the strain but makes the analysis robust to unknown contributions within a single pulsar. We observe a single pulsar with $ \Delta \dot{ P} _b $ away from zero at above two standard deviations, PSR J2129-5721, with a significance of $ 2.1 \sigma $ (the mean expected number of such pulsars for a $ 14 $ pulsar set is $ 0.6 $). 
\begin{table}
\begin{center}
\caption{Pulsars used for $\dot{P}_b$ analysis. $( l, b ) $ is the galactic longitude and latitude, $d _a $ is the distance between the Earth and the pulsar ($ a $), $T$ is the observation time, $\dot{P}_{b,\text{obs}}/P_b$ is the observed value of the line-of-sight acceleration, $ \dot{P}_{b,\text{int}}/ P_b$ is the intrinsic relative change in the binary period induced by gravitational emission, $v _{ \perp } ^2  / d _L $ is the estimated contribution from the Shklovskii effect, $a_{\text{MW}}$ is the estimated contribution from Galactic accelerations (taken with 20\% error bars), $\Delta \dot{ P} _b / P _b $ is the leftover contribution to the orbital pulsar derivative when the prior three are subtracted from $a_{\text{obs}}$, and Ref. is the reference with pulsar parameters from which these accelerations can be computed. All contributions to $ \dot{P}_{b}/P_b $ listed below are in units of $10^{-18}$ s$^{-1}$. Pulsars for which the intrinsic contribution has been estimated using the quadrupole approximation to binary radiation are demarcated with a $\dagger$ (*) with inputs taken from Ref.~\cite{Reardon_2021} (\cite{Fonseca_2016}). Intrinsic entries denoted by `0(0)' do not have individually well-measured binary masses, however we have confirmed these the intrinsic values are negligible for all pulsar masses below the maximum known value, $2.1~M_\odot$.}
\label{tab:Pbdot}
\begin{tabular}{lllllllllll}
\toprule
Pulsar &    $l$ (deg) &     $b$ (deg) &          $d _a $ (kpc) &     $T$ (yr) &       $\dot{P}_{b,\text{obs}}/P_b$ &           $  \dot{P}_{b,\text{int}}/ P_b$ &      $ v _{ \perp  } ^2  / d _L  $ &      $a_\text{MW}$ &    $ \Delta \dot{ P} _b / P _b $  & Ref. \\
\midrule
J0437-4715   &  253.39 & -41.96 &  0.1570(22) &   4.76 &    7.533(12) &   -0.00552(10) &   7.59(11) &   -0.055(11) &   0.0(1) &             \cite{10.1093/mnras/stv2395} \\
J0613-0200   &  210.41 &  -4.10 &     0.80(8) &  16.10 &     0.46(11) &      -$0.02(5)^\dagger$ &  0.215(22) &     0.046(9) &    0.2(1) &              \cite{10.1093/mnras/stw483} \\
J0737-3039AB &  245.24 &  -4.50 &    1.15(18) &   2.67 &  -142.0(1.9) &   -141.565(15) &  0.053(16) &   -0.056(11) &   -0.5(19) &       \cite{doi:10.1126/science.1132305} \\
J0751+1807   &  202.73 &  21.09 &    1.22(25) &  17.60 &    -1.54(11) &      -1.91(17) &   0.56(12) &    0.048(10) &   -0.24(23) &              \cite{10.1093/mnras/stw483} \\
J1012+5307   &  160.35 &  50.86 &    1.41(34) &  16.80 &      1.17(8) &    -0.1955(33) &     2.3(5) &   -0.070(14) &     -0.8(5) &              \cite{10.1093/mnras/stw483} \\
J1022+1001   &  231.79 &  51.50 &   0.719(21) &   5.89 &     0.82(34) &    -0.0021(19) &  0.512(15) &   -0.130(26) &    0.4(3) &             \cite{10.1093/mnras/stv2395} \\
J1537+1155   &   19.85 &  48.34 &    1.16(24) &  22.00 &    -3.766(8) &     -5.3060(8) &     1.8(4) &     -0.19(4) &     -0.09(38) &                      \cite{Fonseca_2014} \\
J1603-7202   &  316.63 & -14.50 &      0.9(7) &   6.00 &     0.57(28) &         0.0(0) &   0.13(10) &    -0.039(8) &    0.5(3) &             \cite{10.1093/mnras/stv2395} \\
J1614-2230   &  352.64 &   2.19 &     0.65(4) &   8.80 &     2.10(17) &   -0.000558(5)* &   1.66(12) &    0.079(16) &    0.4(2) &                         \cite{Alam_2020} \\
J1713+0747   &   28.75 &  25.22 &     1.15(5) &  21.00 &    0.058(26) &  -1.03(6)e-06$^\dagger$ &   0.111(5) &   -0.060(12) &  0.007(28) &             \cite{10.1093/mnras/sty2905} \\
J1738+0333   &   27.72 &  17.74 &    1.47(11) &  10.00 &    -0.56(10) &       -0.91(6) &  0.270(20) &  -0.0049(10) &    0.09(12) &  \cite{10.1111/j.1365-2966.2012.21253.x} \\
J1909-3744   &  359.73 & -19.60 &   1.161(18) &  15.00 &   3.8645(10) &   -0.02111(23) &    3.88(6) &     0.034(7) &     -0.02(6) &                          \cite{Liu_2020} \\
J2129-5721   &  338.01 & -43.57 &    0.53(25) &   5.87 &       1.4(6) &         0.0(0) &    0.19(9) &   -0.121(24) &      1.3(6) &             \cite{10.1093/mnras/stv2395} \\
J2222-0137   &   62.02 & -46.08 &  0.2672(11) &   4.00 &       0.9(4) &    -0.0365(19) &   1.324(5) &   -0.098(20) &     -0.2(4) &                      \cite{Cognard_2017} \\
\bottomrule
\end{tabular}
\end{center}
\end{table}

We show the data for the pulsars used in the $ \ddot{ P } $ analysis in Tab.~\ref{tab:Pddot}. In addition to the longitude, latitude, and distance to the pulsar, we provide the $ \ddot{ P } _{ {\rm obs}}/ P $ as calculated in Ref.~\cite{Liu:2019iuh} using data from Refs.~\cite{10.1093/mnras/stw483,10.1093/mnras/stv2395}. Since there are no known contributions to $ \ddot{ P} _{ {\rm obs}} /P $ at the level of the observed uncertainties (see App.~\ref{app:Pdd} for details), we only report the observed value. The data set contained two pulsars with known contributions to their second derivatives: PSR J1024-0719 and PSR B1821-24A. PSR J1024-0719 is believed to be in a wide binary orbit with a period between 2000-20000 years that is inducing a large $ \ddot{ P}  _{ {\rm obs}}$~\cite{Kaplan:2016ymq}, while PSR B1821-24A is known to be in a dense cluster where accelerations are significant~\cite{Bilous_2015}, and hence we exclude both these pulsars from our study. The dataset also includes a pulsar with a $ \ddot{ P } _{ {\rm obs}} $ significantly away from zero, PSR B1937+21, that requires more scrutiny. The pulsar does not appear to be in a binary orbit nor in a dense cluster but still appears to have an unusually large value of $ \ddot{ P} _{ {\rm obs}} /P $~\cite{Liu:2019iuh}. Given the measurements of other pulsars in the dataset, this value cannot be explained by a background gravitational wave. There has been some discussion in the literature if its $ \ddot{ P} $ value might be due to mismodeled red noise~\cite{1994ApJ...428..713K}. Instead of incorporating PSR B1937+21 in our analysis, we take it as an outlier and omit it from our analysis. Given the sensitivity of PSR B1937+21 relative to the other pulsars in our set, this does not significantly degrade our sensitivity.

After removing the three suspect pulsars, there are two pulsars with $ \ddot{ P} _{ {\rm obs}}  $ discrepant from zero at greater than two standard deviations: PSR J0621+1002 ($ 2.2 \sigma $) and PSR J1022+1001 ($ 2.3 \sigma $). This is within the expected level of statistical fluctuations; for a 46-pulsar dataset, the mean number of pulsars with fluctuations at this level is 2.1. 

\begin{table}
\begin{center}
\caption{Pulsars used for $\ddot{P}$ analysis. $l$ is galactic longitude, $b$ is galactic latitude, $d$ is the distance between the Earth and the pulsar, $T$ is the observation time, $\ddot{P}_\text{obs}/P $ is the observed line-of-sight jerk, $\sigma_\text{RN}$ is the additional uncertainty on $ j _{ {\rm GW}} $ due to ultralow-frequency red noise, and Ref. is the reference with pulsar parameters from which these jerks can be computed. The three pulsars at the bottom of the table were not used to in the gravitational wave search (see main text for further details).} 
\label{tab:Pddot}
\setlength{\tabcolsep}{0.5em}
\bgroup
\def\arraystretch{.75}%
\begin{tabular}{llllllll}
\toprule
Pulsar &       $ l$ (deg) &       $b$ (deg) &      $d$ (kpc) &    $T$ (yr) &        $\ddot{P}_\text{obs}/P$ ($10^{-30}$ s$^{-2}$) &  $\sigma_\text{RN}$  ($10^{-30}$ s$^{-2}$) & Ref. \\
\midrule
J0030+0451 &  113.141 & -57.611 &  0.324 &  15.1 &          -4(4) &    0.54       &  \cite{10.1093/mnras/stw483} \\
J0034-0534 &  111.492 & -68.069 &  1.348 &  13.5 &          0(20) &           0 & \cite{10.1093/mnras/stw483} \\
J0218+4232 &  139.508 & -17.527 &  3.150 &  17.6 &          -2(5) &            0.14& \cite{10.1093/mnras/stw483} \\
J0437-4715 &  253.394 & -41.963 &  0.157 &  14.9 &      -1(1) &            0  & \cite{10.1093/mnras/stv2395} \\
J0610-2100 &  227.747 & -18.184 &  3.260 &   6.9 &       0(50) &           0 & \cite{10.1093/mnras/stw483} \\
J0613-0200 &  210.413 &  -9.305 &  0.780 &  16.1 &         0.6(6) &         0.13   & \cite{10.1093/mnras/stw483} \\
J0621+1002 &  200.570 &  -2.013 &  0.425 &  11.8 &        -70(30) &         1.28   & \cite{10.1093/mnras/stw483} \\
J0711-6830 &  279.531 & -23.280 &  0.106 &  17.1 &       1(1) &             0 & \cite{10.1093/mnras/stv2395} \\
J0751+1807 &  202.730 &  21.086 &  1.110 &  17.6 &       0(2) &            0 & \cite{10.1093/mnras/stw483} \\
J0900-3144 &  256.162 &   9.486 &  0.890 &   6.9 &        -10(20) &          0  & \cite{10.1093/mnras/stw483} \\
J1012+5307 &  160.347 &  50.858 &  0.700 &  16.8 &         0.4(7) &          0.01  & \cite{10.1093/mnras/stw483} \\
J1022+1001 &  231.795 &  51.101 &  0.645 &  17.5 &      -2(1) &           0.01  & \cite{10.1093/mnras/stw483} \\
J1045-4509 &  280.851 &  12.254 &  0.340 &  17.0 &          -2(7) &          0.05   & \cite{10.1093/mnras/stv2395} \\
J1455-3330 &  330.722 &  22.562 &  0.684 &   9.2 &          6(20) &            0.17 & \cite{10.1093/mnras/stw483} \\
J1600-3053 &  344.090 &  16.451 &  1.887 &   9.1 &           4(5) &             0.06 & \cite{10.1093/mnras/stv2395} \\
J1603-7202 &  316.630 & -14.496 &  0.530 &  15.3 &           1(4) &             0.02 & \cite{10.1093/mnras/stv2395} \\
J1640+2224 &   41.051 &  38.271 &  1.515 &  17.3 &        -0.9(9) &            0 & \cite{10.1093/mnras/stw483} \\
J1643-1224 &    5.669 &  21.218 &  0.740 &  17.3 &      -2(2) &           0.01  & \cite{10.1093/mnras/stw483} \\
J1713+0747 &   28.751 &  25.223 &  1.311 &  17.7 &        -0.5(5) &      0.38    &   \cite{10.1093/mnras/stw483} \\
J1721-2457 &    0.387 &   6.751 &  1.393 &  12.7 &      -30(70) &           0.01  & \cite{10.1093/mnras/stw483} \\
J1730-2304 &    3.137 &   6.023 &  0.620 &  16.9 &       0(2) &            0  & \cite{10.1093/mnras/stv2395} \\
J1732-5049 &  340.029 &  -9.454 &  1.873 &   8.0 &         20(20) &        0.03  &    \cite{10.1093/mnras/stv2395} \\
J1738+0333 &   27.721 &  17.742 &  1.471 &   7.3 &      -30(90) &          0  & \cite{10.1093/mnras/stw483} \\
J1744-1134 &   14.794 &   9.180 &  0.395 &  17.3 &         0.8(8) &          0.02  & \cite{10.1093/mnras/stw483} \\
J1751-2857 &    0.646 &  -1.124 &  1.087 &   8.3 &      -10(50) &           0 & \cite{10.1093/mnras/stw483} \\
J1801-1417 &   14.546 &   4.162 &  1.105 &   7.1 &  -30(100) &           0.02 & \cite{10.1093/mnras/stw483} \\
J1802-2124 &    8.382 &   0.611 &  0.760 &   7.2 &       10(60) &           0.01 & \cite{10.1093/mnras/stw483} \\
J1804-2717 &    3.505 &  -2.736 &  0.805 &   8.1 &      -40(40) &           0 & \cite{10.1093/mnras/stw483} \\
J1843-1113 &   22.055 &  -3.397 &  1.260 &  10.1 &         -7(20) &        0.05  &   \cite{10.1093/mnras/stw483} \\
J1853+1303 &   44.875 &   5.367 &  2.083 &   8.4 &        -30(20) &         0  &  \cite{10.1093/mnras/stw483} \\
B1855+09   &   42.290 &   3.060 &  1.200 &  17.3 &       1(2) &           0.03  & \cite{10.1093/mnras/stw483} \\
J1909-3744 &  359.731 & -19.596 &  1.140 &   9.4 &         0.6(9) &     0.02     &   \cite{10.1093/mnras/stw483} \\
J1910+1256 &   46.564 &   1.795 &  1.496 &   8.5 &         30(20) &          0  & \cite{10.1093/mnras/stw483} \\
J1911+1347 &   25.137 &  -9.579 &  1.069 &   7.5 &          14(8) &           0  & \cite{10.1093/mnras/stw483} \\
J1911-1114 &   47.518 &   1.809 &  1.365 &   8.8 &       20(50) &          0   & \cite{10.1093/mnras/stw483} \\
J1918-0642 &   30.027 &  -9.123 &  1.111 &  12.8 &           0(8) &        2.46     & \cite{10.1093/mnras/stw483} \\
B1953+29   &   65.839 &   0.443 &  6.304 &   8.1 &      -20(50) &       0      & \cite{10.1093/mnras/stw483} \\
J2010-1323 &   29.446 & -23.540 &  2.439 &   7.4 &         20(20) &        0    & \cite{10.1093/mnras/stw483} \\
J2019+2425 &   64.746 &  -6.624 &  1.163 &   9.1 &      -500(900) &      0     &  \cite{10.1093/mnras/stw483} \\
J2033+1734 &   60.857 & -13.154 &  1.740 &   7.9 &  40(100) &         0    & \cite{10.1093/mnras/stw483} \\
J2124-3358 &   10.925 & -45.438 &  0.410 &  16.8 &       0(3) &         0.02     & \cite{10.1093/mnras/stv2395} \\
J2129-5721 &  338.005 & -43.570 &  3.200 &  15.4 &      -1(2) &         0    &  \cite{10.1093/mnras/stv2395} \\
J2145-0750 &   47.777 & -42.084 &  0.714 &  17.5 &      -2(1) &        0.28    & \cite{10.1093/mnras/stw483} \\
J2229+2643 &   87.693 & -26.284 &  1.800 &   8.2 &        -20(20) &    0    &     \cite{10.1093/mnras/stw483} \\
J2317+1439 &   91.361 & -42.360 &  1.667 &  17.3 &      -1(3) &          0   & \cite{10.1093/mnras/stw483} \\
J2322+2057 &   96.515 & -37.310 &  1.011 &   7.9 &       30(70) &        0    & \cite{10.1093/mnras/stw483} \\
\bottomrule
\end{tabular}
\egroup
\end{center}
\end{table}

\subsection{Statistical Analysis}
\label{sec:statistics}
In order to compute a limit, we work with a test statistic $\hat{q}$ that we define via a log-likelihood ratio. We begin by defining a likelihood $\mathcal{L}$ as a multivariate Gaussian over the dataset under the assumption of uncorrelated uncertainties between pulsars:
\begin{equation}
{\cal L} (h_0, f_\text{GW},{\bm \theta} | \left\{ y _a  \right\} ) =
\prod_{a=1}^{N_\text{p}} \frac{1}{\sqrt{2\pi}\sigma_a} \exp\left[-\frac{(y _a  - \bar{y} _a   (h_0, f_\text{GW},{\bm \theta} ) )^2}{2\sigma_a^2}\right]\,.
\label{eq:likelihood}
\end{equation}
Here, we have denoted the measured data (either {$j_{\text{GW},a}$ or $a_{\text{GW},a}$) as $ y _a $ with associated uncertainty $ \sigma _a $, where $a$ indexes the $N_\text{p}$ pulsars in the dataset~\footnote{The uncertainty, $ \sigma _a $, is the quadrature sum of uncertainties associated with the observed value, any deterministic contributions, and our estimate for ultralow-frequency red noise.}. The model prediction, $\bar{y}_a( h_0, f_\text{GW},{\bm \theta} )$, is given by the time derivatives of the line-of-sight velocity induced by a GW, Eq. (1) in the main text. We adopt an instantaneous approximation to the time derivatives, hence take $\bar{y} _a   (h_0, f_\text{GW},{\bm \theta} ) = (\pi f_\text{GW})^n\,v_{\text{GW}, a}(h_0, f_\text{GW},{\bm \theta} )$ where $n = 1\,(2)$ for acceleration (jerk). In both cases, the model contains seven independent parameters, given by $f_\text{GW}, h_0$, and five angular parameters corresponding to sky position and orbital orientation, which we group into a set of nuisance parameters $ {\bm \theta} \equiv  (l, b, i,\psi, \Phi_0) $. Note that these parameters refer to the \textit{source} binary black hole inspiral, not the pulsar system. The coordinates $l$ and $b$ are the Galactic longitude and latitude of the GW source, while $i, \psi,$ and $\Phi_0$ are orbital parameters of the SMBH defined in Sec.~\ref{app:SMBHform}.

In order to place a limit, we marginalize over these nuisance parameters by integrating the likelihood over a prior distribution on the nuisance parameters, denoted $\pi({\bm \theta})$, that is uniform on the celestial sphere and over orientations of the SMBH binary:
{
\arraycolsep=1.4pt\def\arraystretch{2.2}
\begin{equation}
\begin{array}{lc}\displaystyle\pi(l)  =  \frac{ 1 }{ 2 \pi };~& 0^\circ \leq l < 360^\circ \\ 
\pi(b) =  \displaystyle\frac{1}{2} \sin(b);~& 0^\circ \leq b < 180^\circ\\
\pi(i)  = \displaystyle\frac{1}{2} \cos(i);~&-\pi/2 < i < \pi/2\\
\pi(\psi)  =  \displaystyle\frac{ 1 }{ 2\pi };~& -\pi/2 < \psi < 3\pi/2\\
\pi(\Phi_0)  =  \displaystyle\frac{ 1 }{ 2\pi };~& 0 <\Phi_0 < 2\pi
\end{array}\,.
\end{equation}
}

We compute the marginalized likelihood,
\begin{equation} 
\mathcal{L}_\text{M}(h_0, f_\text{GW}) \equiv \int \mathcal{L}(h_0, f_\text{GW},{\bm \theta}  ) \pi({\bm \theta})\, d{\bm \theta}\,,
\end{equation} 
using Monte Carlo integration with $10 ^3 $ samples. At a given fixed frequency $f_\text{GW}$, we introduce the test statistic using the marginalized log-likelihood ratio,
\begin{equation} 
\hat{q} ( h _0)  = 2 \log \left( \frac{ \mathcal{L} _{ {\rm M}} ( h _0  , f _{ {\rm GW}} )  }{ \mathcal{L} _{ {\rm M}} ( 0 ,f _{ {\rm GW}} ) } \right) \,,
\end{equation} 
where the numerator is evaluated at a signal strength, $ h _0 $, and the denominator is evaluated under the null hypothesis of no GW signal ($h_0 = 0$). 

Our goal is to find the value of $ h _0 $ for which $ \hat{q} ( h _0 ) $ is inconsistent with the null distribution. According to Wilks' theorem, the asymptotic probability distribution of $ \hat{q} ( h _0 ^{ \star} ) $ should approach a $ ( \chi ^2 _{ ( 1 ) } +1 ) /2 $ distribution, where $ \chi ^2  _{ ( 1 ) } $ is the chi-squared distribution with a single degree of freedom~\cite{Algeri_2020} and $ h _0 ^{ \star } $ is the best-fit value. In performing a maximum likelihood estimation, the corresponding 95\% confidence interval corresponds to the $ h _0 $ region for which $ \hat{q} ( h _0 ) < 2.71$. 

To check the validity of Wilks' theorem, we study the behavior of the test statistic by creating mock data with a true value of $ h _0 $ set to zero. We generate $ 10 ^3 $ samples of $ \left\{ y_a \right\} $ from a Gaussian distribution with mean zero and standard deviation $\left\{ \sigma_a \right\} $. We study the resulting $\hat{q} ( h _0 ^{ \star}) $ null distribution. The results are shown in Fig.~\ref{fig:nulldist} at a reference frequency of $f_\text{GW} = 10^{-10}~{\rm Hz} $ over the pulsars in the $j_\text{GW}$ dataset. The 95th percentile of this particular distribution is $\hat{q} ( h _0 ^{\star})  = 2.68$, which coincides with the expectation of approximately $2.71$ from Wilks' theorem. While we show this result only for the $j _{ {\rm GW}} $ analysis for a fixed frequency, we confirmed the null distribution maintains this form over the frequency range of our analysis for both the $a_\text{GW}$ and $j_\text{GW}$ datasets. 

\begin{figure}[t]
\begin{center} 
\includegraphics[height=8cm]{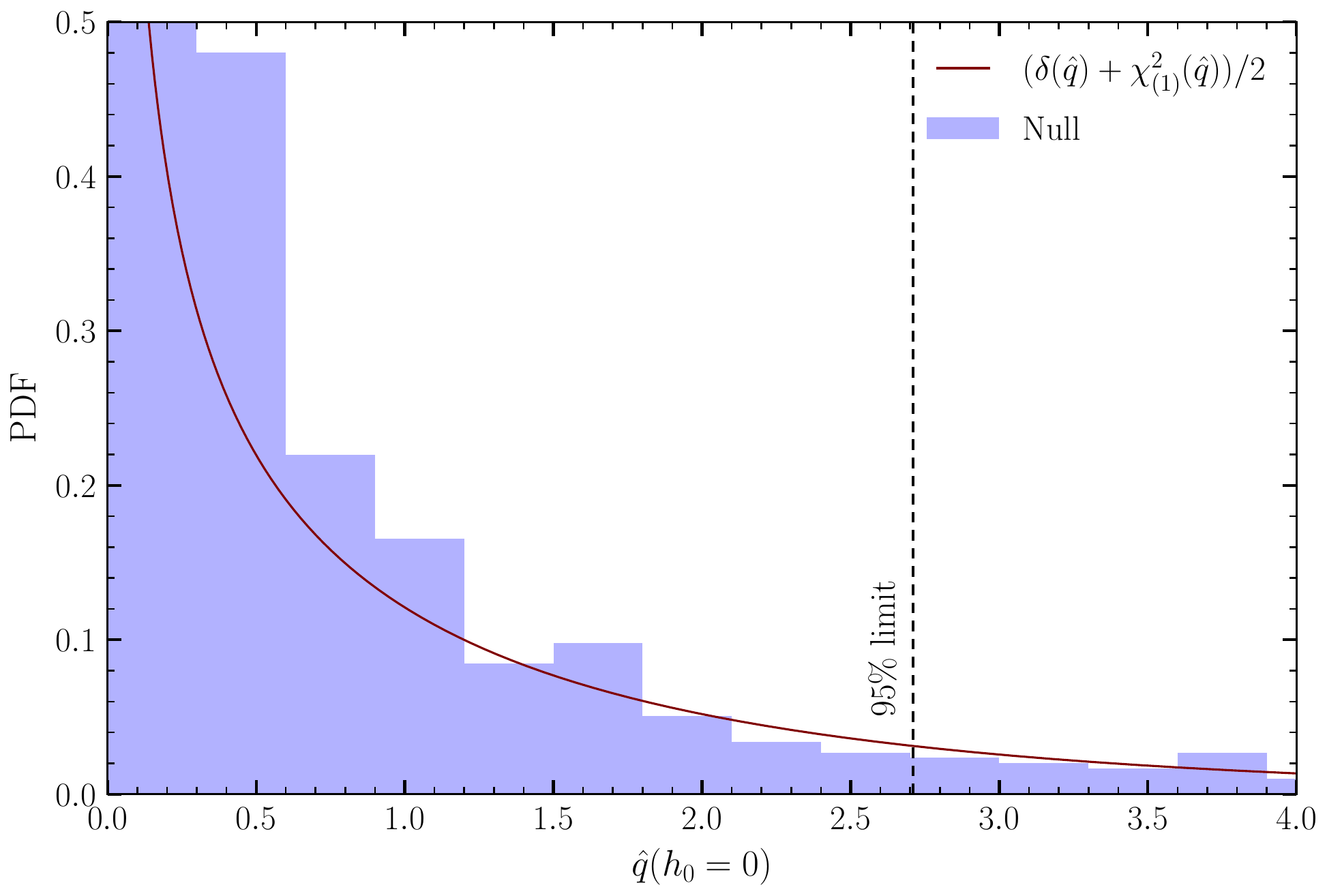}
\end{center}
\vspace{-0.75cm}
\caption{Null distribution of $\hat{q}$ for the pulsars in the $j_\text{GW}$ dataset at a reference frequency of $f_\text{GW} = 10^{-10}$ Hz ({\bf \color{myblue} blue}) compared to the analytic prediction from Wilk's theorem, $(\delta ( \hat{q} ) + \chi^2_{(1)} (\hat{ q} ) ) /2 $ ({\bf \color{Ourcolor} red}). The ninety-fifth percentile of this distribution lies at $\hat{q} = 2.68$, which agrees well with the analytic prediction of 2.71~\cite{Algeri_2020}.}
\label{fig:nulldist}
\end{figure}

Given the excellent agreement between the expected and analytic null distributions, we use the analytic distribution with a detection threshold of $ 95\% $ to search for a signal in the data. Performing this analysis only once introduces numerical effects owing to the Monte Carlo integration procedure. To reduce numerical fluctuations, we perform an identical analysis 100 times at each frequency with different random samples of the nuisance parameters used in the integration. We use the likelihood to calculate the best-fit $ h _0 $ for each sample and take the median of the 100 samples as our estimate of $ h _0 ^{ \star } $. The results are shown in the top panel of Fig.~\ref{fig:res} for $ \ddot{ P} $ ({\bf Left}) and $ \dot{ P} _b $ ({\bf Right}) analysis. Furthermore, we compute the $ p $-value corresponding to the probability of finding a best-fit $ h _0 $ as at least as large as observed, given the null distribution. For the $ \ddot{ P} $ search, we find a minimum $ p $-value of $ 0.21 $, while for the $ \dot{ P} _b $ search, we find a minimum $ p $-value of $ 0.38 $. Neither of these values drops below our threshold of $ 0.05 $; we conclude there is no significant evidence for a signal in the data.

Given the lack of observed signal, we set a limit on the GW amplitude for each sample by finding the $ h _0 $ value for which $\hat{q}(h_0) =  2.71$. Since the log-likelihood ratio appears to obey Wilks' theorem, this threshold corresponds to a 95\% confidence limit. At each frequency, we take the median of the limits produced by each of the 100 realizations, which gives the limits shown in Fig.~\ref{fig:sensitivity} of the main text.

\begin{figure} 
\includegraphics[width=8.5cm]{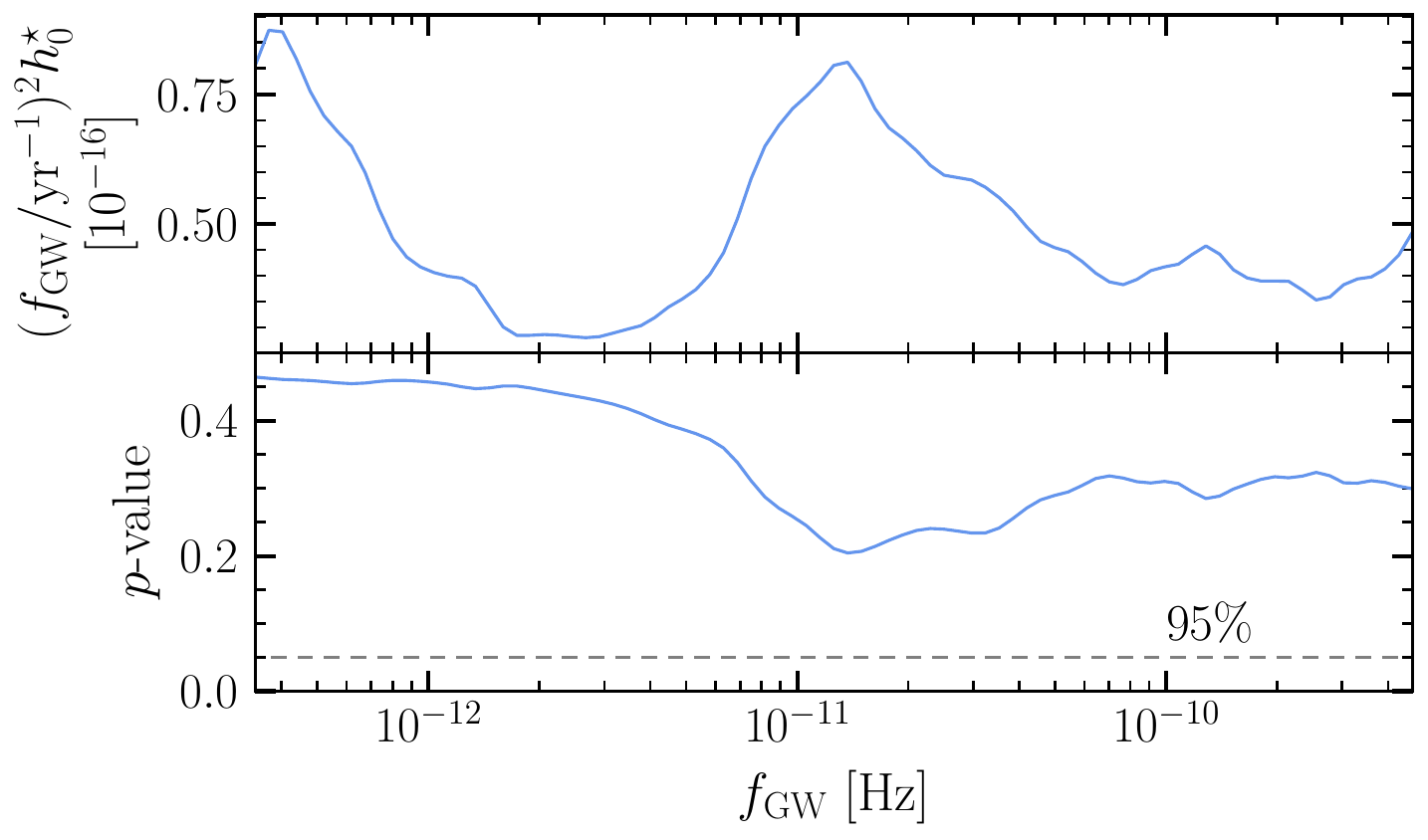}
\includegraphics[width=8.4cm]{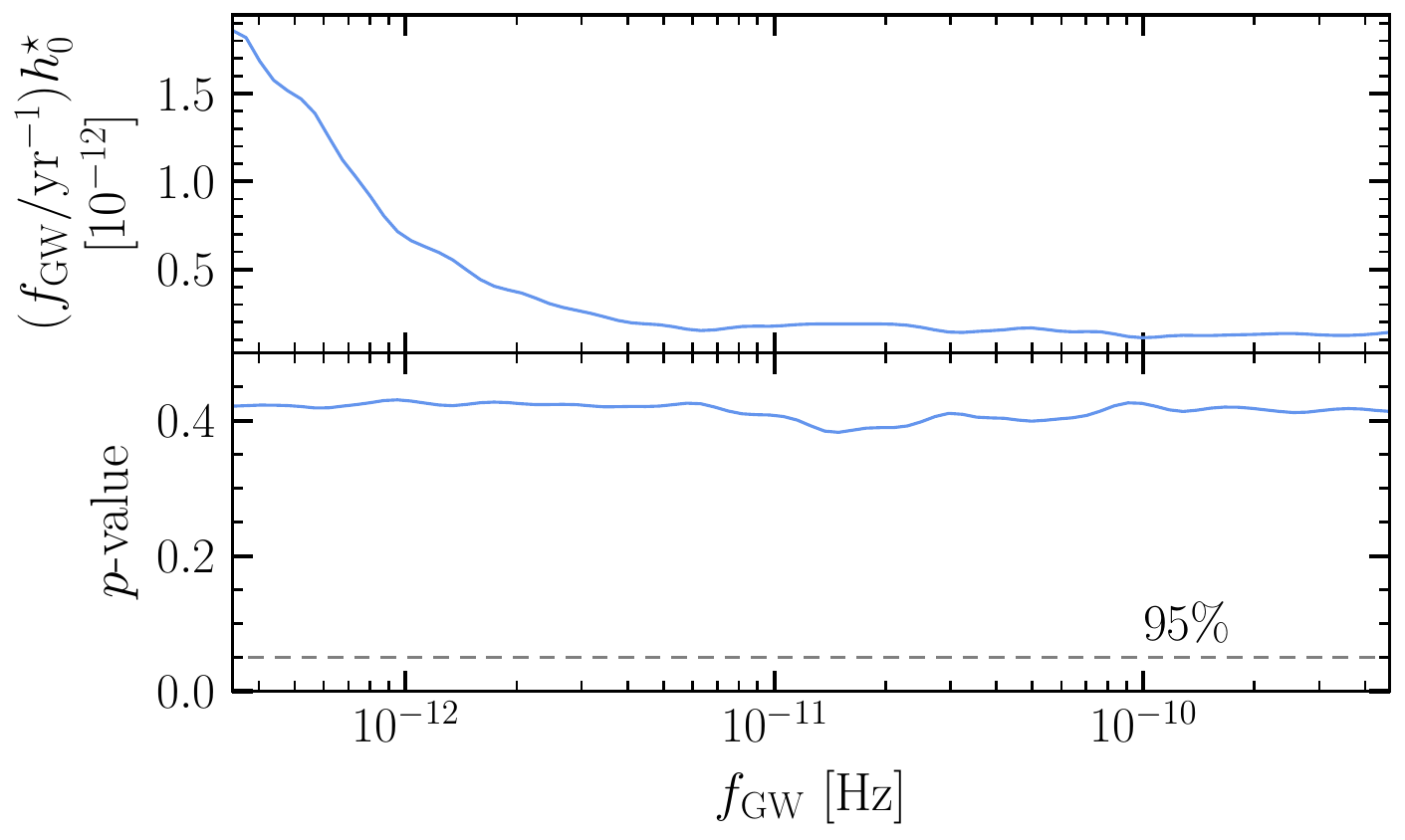}
\caption{Results of searches for a GW signal using the $ \ddot{ P} $ ({\bf Left}) and $ \dot{ P } _b $ ({\bf Right}) analyses. In the top panels, we show the best-fit values $h_0^\star$ multiplied by the expected frequency scaling in the regime, $ d ^{-1} \ll f _{ {\rm GW}} \ll T ^{-1}$. In the bottom panel, we show the corresponding $p$-values of the maximum-likelihood value being as large as observed given the null distribution. We do not observe a statistically significant signal in either search.} 
\label{fig:res}
\end{figure}

\section{GW-Induced Acceleration and Jerk}
\label{app:derivation}
Our results all follow from the expression for the relative SSB-pulsar velocity ($ v _{ {\rm GW}} $) induced by a GW (Eq.~\eqref{eq:vEP}). For completeness, we derive this result from first principles. We refer the reader to Ref.~\cite{Maggiore:2018sht} for additional details. 

To derive $v  _{ {\rm GW}} ( t ) $, we work in the transverse-traceless frame where coordinates are fixed as the GW passes through the system and the observed time of arrival of a pulse ($ t _{ {\rm obs}} $) is emitted at time $ t _e $. The metric is,
\begin{equation} 
d s ^2 =  - d t ^2 + \left[ \delta _{ij} + h _{ij} ( t , {\mathbf{x}} ) \right] d x ^i d x ^j  \,. 
\end{equation} 
The geodesic is determined by setting $ d s ^2 = 0 $, which gives the differential distance. In the transverse-traceless frame, spatial coordinates are held fixed such that the integral over this distance equals the SSB-pulsar distance ($ d _a $). For an incoming light pulse along the $ x $-direction we have, 
\begin{equation} 
  d _a = t _{ {\rm obs}}  - t _{ {\rm  e}} - \frac{1}{2} \int _{ t _{ {\rm e}}} ^{ t _{ {\rm obs}}} d t ' ~h _{ x x } ( t ' , {\mathbf{x}} ( t ' ) ) \,.\label{eq:dpulsar}
\end{equation} 
Since $ h _{ x x  } $ is small, we can drop higher order terms within the $ {\mathbf{x}} ( t ' ) $ argument in $ h _{ x x } $ and set it equal to the unperturbed path, $ {\mathbf{x}} ( t ' ) = ( t _{ {\rm obs}} - t ')  \hat{n} _a $. Furthermore, we can extend Eq.~\eqref{eq:dpulsar} to a generic direction,
\begin{equation} 
t _{ {\rm obs}} =  t _{ {\rm e}} + d _a  + \frac{ n ^i _a n ^j _a }{ 2} \int _{ t _{ {\rm e}}} ^{ t _{ {\rm obs}}} d  t' ~h _{ ij  } ( t ' , ( t _{ {\rm e}} + d _a - t '  ) \hat{n} _a )\,.
\end{equation}  
The observed difference of time-of-arrivals of successive pulses is related to the $ v _{ {\rm GW}} ( t ) $ by the Doppler shift formula, such that,
\begin{equation} 
v _{ {\rm GW}}  ( t ) \simeq \frac{1}{2} n _a ^i n _a ^j \int  _{t_ {\rm e}} ^{ t _{ {\rm e}} + d _a } d t ' \left[ \frac{ \partial }{ \partial t '} h _{ij} ( t ' , {\mathbf{x}} ) \right] _{ {\mathbf{x}} = {\mathbf{x}} _0 ( t ' ) }\,,
\end{equation} 
where $ {\mathbf{x}} _0 ( t ' ) \equiv    ( t _{ {\rm e}} + d _a - t '  ) \hat{n} _a $. For a monochromatic GW propagating along the $ \hat{n} $ direction,
\begin{equation} 
h _{ij} ( t , {\mathbf{x}} )  = {\cal A} _{ij} ( \hat{n} ) \cos \left[ \omega  ( t - \hat{n} \cdot {\bf x}   ) \right] \,,
\end{equation} 
we have, 
\begin{equation} 
 v _{ {\rm GW}} ( t ) = \sum _{ A = + , \times } F _a  ^A ( \hat{n} ) \left[ h _{A} ( t ,{\mathbf{x}} = 0 ) - h _{A} ( t - d  _a  , {\mathbf{x}} _a ) \right]\,.
\end{equation} 
Here we set the SSB at the origin and substituted $t _{ {\rm obs}} $ with $ t $. The first term is the impact of the GW on the SSB, and the latter is on the pulsar. This is Eq.~\eqref{eq:vEP} in the main text. 

\section{Kinematic and Galactic Contributions to $ \ddot{ P} $}
\label{app:Pdd}
The observed second derivative of the period has intrinsic, kinematic, and galactic corrections to Eq.~\eqref{eq:Pdd} of the main text, neglected in this work. In this section, we calculate each of these and argue they are small relative to the uncertainty in the observations. 

To calculate the second derivative, we follow the formalism used in Ref.~\cite{10.1093/mnras/sty1202}. The detected period of the pulsar is related to the true period ($ P _{0} $), the relative position of the SSB-pulsar system, $ {\bf r} $, and its time derivative, $ \dot{ {\bf r}} $, through the Doppler shift,
\begin{equation} 
P = \frac{ P _{ 0}}{1 - \dot{ {\bf r} } \cdot \hat{\bf r}  } \,. \label{eq:Doppler}
\end{equation} 
To derive the observed $ \dot{ P}$ and $ \ddot{ P } $, we first take derivatives of Eq.~\eqref{eq:Doppler}:
\begin{align} 
\dot{ P} & = \frac{ 1 }{ 1 - \dot{{\bf r} } \cdot \hat{\bf r} } \left\{ \dot{ P} _0 + \frac{ P _0 }{  1 - \dot{ {\bf r} } \cdot \hat{\bf r}  } ( \ddot{ {\bf r}} \cdot \hat{\bf r} + \dot{ {\bf r} } \cdot \dot{ \hat{\bf r} } ) \right\} \label{eq:Pdot} \,,\\ 
\ddot{ P} & = \frac{1}{ 1 - \dot{ \hat{{\bf r}} } \cdot \hat{{\bf r}} }  \left\{ \ddot{ P _0 } + \frac{ 2 \dot{ P} _0 }{  1 - \dot{ {{\bf r}} } \cdot \hat{{\bf r}}  } ( \ddot{ {\bf r} } \cdot \hat{{\bf r}} + \dot{ {\bf r} } \cdot \dot{ \hat{{\bf r}} } ) +  \frac{ 2P _0 }{ ( 1 - \dot{ {\bf r} } \cdot \hat{\bf r} ) ^2 } ( \ddot{ {\bf r} } \cdot \hat{\bf r} + \dot{ {\bf r} } \cdot \dot{ \hat{{\bf r}} } ) ^2 + \frac{ P _0 }{ 1 - \dot{ {\bf r} } \cdot \hat{\bf r}}   ( \dddot{ {\bf r} } \cdot \hat{{\bf r}} + 2\ddot{ {\bf r} } \cdot \dot{ \hat{{\bf r}} }  + \dot{ {{\bf r}} } \cdot \ddot{ \hat{{\bf r}} } ) \right\} \label{eq:Pddot}\,.
\end{align} 
$ \dot{ P}  _{ {\rm obs}}$ and $ \ddot{ P} _{ {\rm obs}}$ are given by evaluating Eqs.~\eqref{eq:Pdot} and ~\eqref{eq:Pddot} at the reference time, $ t = 0 $. To this end, we assume changes in the parameters and positions are small such that we can expand $ P _0 = P _{ {\rm int}} + \dot{ P } _{ {\rm int}} t + \frac{1}{2} \ddot{ P} _{ {\rm int}}  t ^2 $ and $ {\bf r} = {\bf d}  + {\mathbf{v}} t + \frac{1}{2} {\bf a } t ^2 $. Using, $ \hat{\bf r} = {\bf r} / | {\bf r} | $ and working to second order in time,
\begin{equation} 
\hat{{\bf r}}  = \hat{\bf d }  + \frac{ {\mathbf{v}} _\perp }{ d  } t +  \frac{1}{2} \left( \frac{ {\bf a } _\perp }{ d  }  - \hat{{\bf d }} \frac{ v _\perp ^2 }{ d  ^2 } - \frac{ 2 v _{ \parallel } {\mathbf{v}} _\perp }{ d  ^2 } \right) t ^2 \,.
\end{equation}
Here we introduced the subscript notation, $ {\mathbf{v}} _\parallel \equiv {\mathbf{v}} \cdot \hat{\bf d } $, $ {\mathbf{v}} _\perp \equiv {\mathbf{v}} - {\mathbf{v}} _{ \parallel } $, and similarly for accelerations. Inserting these expressions into Eqs.~\eqref{eq:Pdot} and ~\eqref{eq:Pddot}, and evaluating the final expression at $ t = 0 $ gives,
\begin{align} 
 \frac{ \dot{ P} _{ {\rm obs}}}{P}&   = \frac{ 1 }{ 1 - v _{ \parallel }}  \left\{  \frac{ \dot{ P } _{ {\rm int}} }{ P  }   + \frac{ 1  }{  1 - v _{ \parallel} } \left(  a _{ \parallel} + \frac{  v _\perp ^2 }{ d  } \right) \right\}\,,  \\ 
\frac{ \ddot{ P } _{ {\rm obs}}}{ P} & = \frac{1}{ 1  - v _{ \parallel}} \left\{ \frac{ \ddot{ P} _{{\rm int}} }{ P } + \frac{ 2 \dot{ P}_{ {\rm int}} / P}{ 1 - v _{ \parallel}} \left(  a _{ \parallel} + \frac{ v _\perp ^2 }{d } \right)  + \frac{ 3 v _\perp /d   }{ 1 - v _{\parallel}} \left( a _\perp  - \frac{ v _\parallel v _\perp }{ d    } \right) + \frac{ 2  }{  ( 1 - v _{ \parallel } ) ^2 } \left( a _{ \parallel } + \frac{ v _\perp ^2 }{ d  } \right) ^2  + \frac{ j _{ \parallel } }{ 1 - v _{ \parallel}} \right\} \,, \label{eq:Pddot+full}
\end{align}  
where in the final step we have redefined $ P _{ {\rm int}} \rightarrow P $ to keep with the convention in the main text. 

Our goal is to estimate the size of the galactic and kinematic contributions to $ \ddot{ P }_{ {\rm obs}}/ P $. With current data, the uncertainty of $ \ddot{ P} _{ {\rm obs}} / P $ reaches $ 1 0 ^{ - 30} ~ {\rm sec} ^{ - 2 } $. The $ \ddot{ P} _{ {\rm int}} / P $ term is the intrinsic contribution. This value is expected to be of order $ {\cal O} ( ( \dot{ P} _{ {\rm int}}/ P ) ^2 ) $~\cite{2012hpa..book.....L}, which is typically $ {\cal O} ( 10 ^{ - 35} ~{\rm sec} ^{ - 2} ) $, well below current experimental uncertainties on $ \ddot{ P} _{ {\rm obs}}/P$. The contribution to the jerk from the galactic potential was estimated in Ref.~\cite{10.1093/mnras/sty1202}. They found this is highly subdominant for each pulsar relative to the observed parameter uncertainty. For the rest of the contributions, taking typical values for pulsars in our dataset, $  v _\perp \sim v _\parallel   \sim 100~{\rm km}/{\rm sec} $, $ d  \sim {\rm kpc} $, $ a _{ {\rm MW}} \sim 10 ^{ - 19}~ {\rm sec} ^{-1} $, and $ \dot{ P } _{ {\rm int}} /P  \sim 10 ^{ - 18} ~ {\rm sec} ^{-1} $, we find the leading correction to our approximation that $\ddot{ P } _{ {\rm obs}} /P = 0$ arises from the $ - 3 v _\perp ^2 v _{ \parallel } / ( 1 - v _{ \parallel} ) ^2 d ^2 $ term. For our benchmark values, we find this term is $ {\cal O} ( 10 ^{ - 33} ~ {\rm sec} ^{-1} ) $. This is about two orders of magnitude below our target sensitivity, justifying our decision to neglect these contributions.

While the various kinematic and galactic contributions to $ \ddot{ P} $ are currently negligible, this will not be the case in the future as growing pulsar observation times reduce the uncertainties in $ \ddot{ P} $ to a level at which these effects become detectable. At that point, maximizing the pulsar timing array sensitivity to ultralow-frequency gravitational waves will require modeling or measuring each of the corrections in Eq.~\eqref{eq:Pddot+full}, as was done in the main text for $ \dot{ P} _b  $. We emphasize that this challenge is present whether a search for ultralow-frequency GWs is conducted using the timing model (as done in our work) or using the residuals, but using the timing model provides a clear way to incorporate this effect into the analysis.

In this discussion, we neglected the influence of a background of GWs on Eq.~\eqref{eq:Pddot+full}. GWs will primarily contribute to $ j _{ \parallel } $, as denoted by $ j _{ {\rm GW}} $ in the main text. Additionally, GWs appear as an acceleration and can, in principle, have a secondary contribution from other terms in Eq.~\eqref{eq:Pddot+full}. The term proportional to $ a _\parallel ^2  $ is small as it is $ {\cal O} ( h _0 ^2 ) $. Furthermore, when $ f _{ {\rm GW}} \gg v  d ^{-1} , \dot{ P} _{ {\rm int}}/ P  $ (the case everywhere in our analysis), the remaining contributions are also negligible. 

\section{Impact of Ultralow-Frequency Red Noise}
\label{app:rednoise}
The analysis performed by Liu et al.~\cite{Liu:2019iuh}  from which we extracted our $ \ddot{ P} $ values did not account for ultralow-frequency red noise. To estimate its influence on our measurement, we derive an expression for the variance of the pulsar timing model parameters, $ \dot{ P} _b $ and $ \ddot{ P} $, induced by ultralow-frequency red noise. 

Our starting point is the induced redshift, $ z $, and the residuals are given as an integral over redshift, $ R ( t )  = \int _0 ^t dt z $~\footnote{The definition of redshift is equivalent to $ v _{ {\rm GW}} ( t ) $ in the main text. In this appendix, we confirm to the conventional notation in the literature.}. We assume the red noise is imprinted as a stochastic background such that we can define a redshift power spectrum,
\begin{equation} 
\left\langle \tilde{ z} ( f  ) \tilde{ z}  ^\ast    ( f  ' ) \right\rangle =\delta ( f  - f  ' ) \frac{1}{2} S _{ {\rm RN}} ( f  ) \,,
\end{equation} 
where, 
\begin{equation} 
 \tilde{z} ( f )  \equiv \int _{ - \infty } ^{\infty} dt ~z ( t ) e ^{ -2\pi i f t }
\end{equation} 
and $ S _{ {\rm RN}} ( - f  ) = S _{ {\rm RN}} ( f  ) $. To ensure our definition of $ S _{ {\rm RN}} ( f ) $ matches the existing literature, we compute the residual correlator, 
\begin{align} 
\left\langle R ( t ) R ( t ' ) \right\rangle & = \frac{1}{2} \int _{ - \infty } ^{\infty}  df  \frac{S _{ {\rm RN}} ( f  ) }{ ( 2\pi f )  ^2 } \big( 1 - e ^{ i 2\pi f  t } \big) \big( 1 - e ^{ - i  2\pi f    t ' } \big) \,,\\ 
& =  \int _0 ^{\infty} df  \frac{ S _{ {\rm RN}} ( f  ) }{ ( 2\pi f )  ^2 }\left( 1 - \cos ( 2\pi f  t )  - \cos ( 2\pi f  t ' ) + \cos ( 2\pi f  ( t - t ' ) ) \right) \,.
\end{align} 
In the limit that $ f t \gg 1 $ and $ f t ' \gg 1 $, the non-stationary pieces are small. Furthermore, dropping the constant piece (which corresponds to an unobservable shift in the residuals), we have,
\begin{align} 
\left\langle R ( t ) R ( t ' ) \right\rangle & \simeq   \int _0 ^{\infty} df  \frac{ S _{ {\rm RN}} ( f  ) }{ ( 2\pi f )  ^2 }\cos ( 2\pi f  ( t - t ' ) )  \,.
\end{align} 
This form is commonly found in the PTA literature (see, e.g., Ref.~\cite{Hazboun:2019vhv}). We now estimate the ensemble variance of the contribution of red noise for the two timing model parameters of interest, 
\begin{align} 
\Big\langle \Big( \frac{ \Delta \dot{P} _b   }{ P _b  } \Big) ^2 \Big\rangle &  = \left\langle \dot{ z} ( t ) ^2 \right\rangle =  \int _0 ^{1/4T } d f ~ S _{ {\rm RN}} ( f  )  ( 2\pi f  ) ^2\,, \\ 
\Big\langle \Big( \frac{ \Delta \ddot{P}  }{ P } \Big) ^2 \Big\rangle &  = \left\langle \ddot{ z} ( t ) ^2 \right\rangle =  \int _0 ^{1/4T} d f ~S _{ {\rm RN}} ( f  )  ( 2\pi f  ) ^4\,.
\end{align} 
In the final step, we cut off the integral at $   1/ 4T  $ as, above this frequency, the instantaneous derivative approximation fails. Evaluating these integrals for fit values of the power-law spectrum, one finds the ultralow-frequency contribution to $ \dot{ P} _b $ is negligible when compared with current uncertainties on $ \dot{ P} _{b, {\rm obs}} $. In contrast, the contribution to $ \ddot{ P} $ can be competitive with the uncertainty in $ \ddot{ P} _{ {\rm obs}} $ and hence we take it into account in our analysis, as described Sec.~\ref{sec:statistics}. 

\bibliographystyle{JHEP}
\bibliography{refs}

\end{document}